\documentstyle[aps,12pt,manuscript]{revtex}
\begin{document}

\def\be{\begin{equation}}
\def\ee{\end{equation}}
\def\bea{\begin{eqnarray}}
\def\eea{\end{eqnarray}}

\def\pd{\partial}
\def\a{\alpha}
\def\b{\beta}
\def\g{\gamma}
\def\d{\delta}
\def\m{\mu}
\def\n{\nu}
\def\t{\tau}
\def\l{\lambda}

\def\s{\sigma}
\def\e{\epsilon}
\def\scri{\mathcal{J}}
\def\cM{\mathcal{M}}
\def\tcM{\tilde{\mathcal{M}}}
\def\RR{\mathbb{R}}

\preprint{INJE-TP-98-9}

\title{New gauge bosons and logarithmic corrections in an exact AdS$_3$}

\author{ Jin Young Kim}
\address{Department of Physics, Kunsan National University, 
Kunsan 573-701, Korea}
\author{ H.W. Lee and Y. S. Myung}
\address{Department of Physics, Inje University, Kimhae 621-749, Korea} 

\maketitle

\begin{abstract}
We calculate the 
absorption cross section by studying the spin--dependent 
wave equation in three-dimensional anti-de Sitter space(AdS$_3$).  
Here the AdS/CFT correspondence is used.
It turns out that the new gauge bosons coupled to (2,0) and (0,2) 
operators on the boundary at infinity receive logarithmic corrections.  
This shows that the gauge bosons may play the role of 
singletons in AdS$_5$.
On the other hand, test fields including 
the intermediate scalars($\eta, \xi$) and fixed scalar($\lambda$) 
do not receive any logarithmic correction in the first-order 
approximation.
\end{abstract}

\newpage

\section{Introduction}
Recently there has been great progress in studying the black hole 
physics using string theory and conformal field theory(CFT)
\cite{Mal9711200,Gub9802109,Wit9802150}.
A 5D black hole (M$_5$: D1-D5 brane black hole) contains the BTZ 
black hole in the near-horizon and thus it is very important to study 
the BTZ black hole(AdS$_3$)\cite{Sfe98NPB179,Mal9804085}. 
Actually, a 5D black hole has the 
geometry : AdS$_3$ in the near-horizon(the throat region) but with 
asymptotically flat space. If one starts  with 
M$_5 \times$S$^1 \times $T$^4$ in the ten-dimensional 
type IIB superstring theory, we have AdS$_3 \times$S$^3 \times$T$^4$ in 
the throat region($r_0 \le r \le R, R=\sqrt{r_1 r_5}$) but finds immediatly 
Minkowski space after passing through the remote boundary($r=R$) 
of AdS$_3$. 
In this sense we call it an AdS$_3$ bubble.
Further the AdS/CFT correspondence 
plays a crucial role in calculating the entropy and greybody 
factor(absorption cross section). 
This correspondence implies that 
the ten-dimensional type IIB bulk theory deep in the throat 
((AdS$_3 \times$S$^3)_R \times$T$^4_{r_1^2/r_5^2}$) 
is related to the two-dimensional gauge theory (dual CFT$_2$) on its 
remote boundary.

On the other hand, one found the logarithmic 
corrections to the absorption cross section of minimal scalars 
from an AdS$_5$ bubble scattering\cite{Gub9803023} and 
an AdS$_3$ bubble scattering\cite{Tay9806132}.
Explicitly, these terms can be related to non-conformally 
invariant operators in their gauge theories.
Nowdays it is very important to interpret the CFT results 
in terms of the bulk AdS results. 
Here the bulk AdS means an exact AdS geometry which is an
infinitely long throat without asymptotically flat space.
In the limit of $R\to \infty$, this picture may come out of 
an AdS bubble. 

We propose the AdS/CFT lore that the conformal limit of the gauge 
theory corresponds to scattering in an exact AdS background.
So it appears that the logarithmic term cannot be 
understood from an exact AdS scattering.
However, the 
scattering analysis in an exact AdS is not an easy matter.
This is so because one cannot define ordinary asymptotic states in an 
exact AdS, due to the timelike boundary and the periodicity 
of geodesics.
To avoid the closed timelike curves, one can use the universal cover by 
ignoring the periodicity of time. 
However, the presence of the timelike boundary leads to major 
differences with physics in Minkowski space.
First one does not have a well-posed initial value problem unless 
one puts boundary conditions there.
Hence we always need boundary conditions at infinity.
These are the Dirichlet or Neumann conditions.
But instead of these, one can use the non-normalizable modes 
to obtain the greybody factor\cite{Lee98PRD104013}.
For Minkowski space the boundary conditions can be decomposed 
into ingoing and outgoing waves, which leads to the usual idea of 
particle and S-matrix.
In AdS background spacetime, solutions to the free wave equation can be 
classified into normalizable and non-normalizable modes\cite{Bal99PRD046003}.
The two types of modes are distingished by their asymptotics and so 
the background can be determined from the boundary condition 
at infinity uniquely up to a choice of normalizable 
component\cite{Dou9902022}.
The first(second) correspond to the states of the theory(boundary 
conditions for the field).

A few of recent works shed light on this 
direction\cite{Sus9901079,Bal9902052,Gid9903048,Lee9903054}.
The vacuum correlators 
$\langle {\cal O}(x_1) {\cal O}(x_2) \cdots {\cal O}(x_n) \rangle_{{\rm CFT}_4}$
of CFT$_4$ are expressed as truncated n-point functions
convolved against the non-normalizable modes. 
These can be interpreted as an S-matrix in an exact AdS$_5$ arising 
from a limit of scattering from an AdS$_5$ bubble with asymptotically flat 
space\cite{Bal9902052}.
Furthermore, in the Poincar\'e coordinates to avoid the closed-time 
like curves, the transition amplitude 
between normalizable modes can be constructed to get correlation 
functions of the dual CFT.
Also Giddings showed that a bounadry S-matrix is defined as an overlap of
``in'' and ``out'' states near the timelike boundary of
AdS\cite{Gid9903048}.
This equals the corresponding correlator in the boundary CFT.
On the other hand, two authors(Myung and Lee) showed that the S-matrix 
of an AdS$_3$ bubble in the dilute gas and low energy limits 
can be derived from an exact AdS$_3$ scattering\cite{Lee9903054}.
This was possible by observing two types of potentials and using 
non-normalizable modes.

In this paper we find the logarithmic terms from an 
exact AdS$_3$ scattering of the new gauge bosons with non-zero spin($s=2$).
It is emphasized that these are states in the CFT$_2$ but are 
absent on the supergravity side\cite{Vaf9804172,Boe9806104}.
We introduce the spin-dependent wave equation 
by switching on the AdS/CFT correspondence. 
Here the spin is defined as $s=h_R-h_L$.
Although this equation has an exact solution in the global coordinates, 
it is very difficult to obtain its solution in the Poincar\'e coordinates.
Hence matching procedure is essential for obtaining 
the absorption 
cross sections for the non-zero spin fields in the low temperature limit.
In AdS$_3$ bubble scattering, one needs a self-dual point 
($r=R$) as a matching point to calculate the greybody 
factor. Here we don't need such a self-dual point, as is needed just in the 
AdS$_3$ bubble approach. Instead, we introduce a spin-dependent matching 
point($Y=\sqrt{2s/\omega}, 1<Y<\infty$) for calculation. 
Then the normalizable modes become the relevant one.

The organization of our paper is as follows. In Sec. II we derive 
the spin-dependent wave equation on AdS$_3$ from the AdS/CFT 
correspondence. And then we calculate the zeroth-order absorption cross 
section in the low-temperature limit. 
We compare it with the known results.
We compute the first-order correction 
to the zeroth-order cross section in Sec. III. 
Finally we discuss our result in Sec. IV.

\section{Zeroth-order Matching}
We start from the equation of motion of the field with mass $m$ and spin $s$
in the global coordinates($\tau, \varphi, \rho$)\cite{Mal9804085,Boe9806104},
\be
\left (\Box + {s^2 \over R^2 \sinh^2 \rho} \right ) \psi = m^2  \psi
\label{eq-zeroth}
\ee
with
\be
R^2 m^2 = 2h_R(h_R-1) + 2h_L (h_L-1) - s^2 = (h_R+h_L)
(h_R+h_L-2) ,
\ee
and $s=h_R-h_L$.  Eq.(\ref{eq-zeroth}) comes from the correspondence between 
states in CAdS$_3$(covered AdS$_3$) and in CFT$_2$. 
In other words, the CAdS/CFT correspondence determines the mass and spin 
on AdS$_3$ in terms of the conformal weights ($h_R, h_L$) on the boundary.
Its solution($\psi=e^{-i \omega \tau} e^{i m \varphi} \psi(\rho)$) 
takes the exact form when 
$m=s$, $\omega=-(h_R + h_L)$ or $(h_R + h_L -2)$ as
\be
\psi(\tau, \varphi, \rho) =
{C_1 \over (\cosh \rho)^{h_R+h_L}} 
e^{-is \varphi} e^{+i(h_R+h_L)\tau} 
+ C_2 (\cosh \rho)^{h_R+h_L-2}
e^{-is \varphi} e^{-i(h_R+h_L-2)\tau}.
\label{sol-psi}
\ee
The first corresponds to the normalizable mode and the second to the 
non-normalizable one. 
Although the CFT is well defined on the 
cylindrical boundary expressed in terms of the global coordinates, 
it is important to find a map between the string 
Hilbert space in PAdS(Poincar\'e patch of CAdS) and 
operators of CFT on the plane.  
This is so because the scattering analysis is 
not problematic within PAdS. 
Then the PAdS/CFT duality relates test fields 
within the Poincar\'e patch to conformal operators on the planar boundary.
For this purpose, 
we introduce the Poincar\'e coordinates($y,t,x$)
\begin{eqnarray}
{1 \over y} &=& \cosh \rho \cos \tau + \sinh \rho \cos \varphi  \nonumber\\
t &=& y \cosh \rho \sin \tau 
\label{poincrd} \\
x &=& y \sinh \rho \sin \varphi.  \nonumber
\end{eqnarray}
On the timelike boundary of $y=0$, this change leads to 
a transformation:
$u/v \equiv x \pm t = \pm \tan \left ( {{\tau\pm\varphi} \over 2} \right ) $. 
In terms of these, the three dimensional metric takes
a simple form
\be
{d s^2} = {R^2 \over y^2} (dy^2 + du d v ).
\label{dsl}
\ee
However, the wave equation leads to a complicated form as\cite{Lee9808002}
\be
\left \{ y^2 ( \partial_y^2 - {1 \over y} \partial_y  - \partial_t^2 +
\partial_x^2 ) + { 4 s^2 y^2 \over 4 x^2 + (1 - y^2 + t^2 - x^2)^2
} \right \} \psi = m^2 R^2 \psi.
\label{eq-ytx}
\ee
This is our key equation for studying the exact AdS$_3$ scattering 
for a test field $\psi$ with the non-zero spin.  
It is not easy to find its solution.
In the case of $s=0$, Eq.(\ref{eq-ytx}) can be 
derived from the conventional action on AdS$_3$ as 
\be
I(\psi) = {1 \over 2} \int d^3 x \sqrt{g} \left [ 
  \left ( \nabla \psi \right )^2 + m^2 \psi^2 \right ].
\label{action}
\ee
In order to get the spin-dependent term, we switch on
the CAdS$_3$/CFT$_2$ correspondence as in Eq.(\ref{eq-zeroth}).
As far as we know, there is no way to introduce the spin-dependent 
potential in Eq.(\ref{eq-ytx}) except this method.

Let us study the limiting behavior of the spin-dependent
term. In the limit of $y \to \infty$(infinity), one can approximate
this term to $4 s^2/y^2$ regardless of $x$ and $t$. 
However in the limit of $y \to 0$(horizon), 
we cannot neglect $t$ and $x$ dependence in the denominator, so
we will approximate this term $4 s^2 y^2/R^{\prime 4} $, where $R'$ 
can be considered as the
(compactification) scale along the $x$-direction. 
For a 5D black hole, we have a compactified scale for $x$, but for 
6D black string, one has an uncompactified scale\cite{Sat9810135}.
Keeping this in mind, we consider a plane wave solution 
$\psi(y,t,x) = e^{- i \omega t + i p x} \psi(y)$\cite{Mal9804085}. 
The equation of motion (\ref{eq-ytx}) leads to 
\begin{eqnarray}
y \to \infty &: & \left \{ y^2 ( \partial_y^2 - {1 \over y} \partial_y  
+ \omega^2 - p^2 )
+ { 4 s^2 \over y^2 } \right \} \psi = m^2 R^2 \psi  
\label{eq-infinity}   \\
y \to 0 &: & \left \{ y^2 ( \partial_y^2 - {1 \over y} \partial_y  
+ \omega^2 - p^2 )
+ { 4 s^2 y^2 \over R^{\prime 4} } \right \} \psi = m^2 R^2 \psi.
\label{eq-ytx0}
\end{eqnarray}
Now the spacetime is divided into near and far regions defined
by $y < Y$ and $y > Y $, where $Y$ is some scale to be
determined later. For simplicity we set $p=0$. We also
consider the low energy scattering and so assume $\omega Y <  1$.

In the far region of $y > Y$, we choose a reciprocal variable $z = 2 s /y$
in terms of which the wave function $\psi(y) = \phi (z) / z$
satisfies
\be 
z^2 \phi^{\prime\prime} + z \phi^\prime + (z^2 - \nu^2) \phi
= - { 4 \omega^2 s^2 \over z^2} \phi,
\label{neareq}
\ee
where $\nu^2 = 1 + m^2 R^2$($\nu=h_R+h_L-1$).  
Here we consider only an integer $\nu > 0$. The 
$\nu=0$ case corresponds to the tachyon, which is out of the scope of 
this paper. 
The leading order solution of this equation is
\be
\phi(z) = H^{(2)}_\nu(z),
\ee
where we choose the solution to be purely infalling at the
horizon($y \to \infty$).
We regard the right hand side as a small perturbation in the far
region.

In the near region of $y < Y$, assuming $4 s^2 /R^{\prime 4} \ll \omega^2$,
we introduce a new variable $\sigma = \omega y$
and $\psi(y) = \sigma f(\sigma)$. Then the equation of motion becomes
\be 
\sigma^2 f^{\prime\prime} + \sigma f^\prime + (\sigma^2 - \nu^2) f
= - { 4 s^2 \sigma^2 \over \omega^2 R^{\prime 4}} f.
\label{fareq}
\ee
For small $\sigma$, the right hand side is negligible and the
leading order solution is given by the Bessel functions
\be
f_\nu(\sigma) = \alpha J_\nu (\sigma) + \beta Y_{\nu} (\sigma) ,
\label{sol-fnu}
\ee
where the first term corresponds to normalizable mode, 
whereas the second to non-normalizable mode for $\nu > 1$. 
The normalizable mode is compatible with Dirichlet boundary 
condition at infinity($\sigma=0$), but the non-normalizable mode is not.

As is in the case of AdS bubble\cite{Gub9803023,Tay9806132}, we can
introduce a matching point($Y$) for calculation.
$Y$ is determined by $\omega Y = 2 s
/Y$, i.e. $Y = \sqrt{2s/\omega}$. One finds 
$Y > 1, \omega Y < 1$ in the low energy($\omega < 1$). 
We note that this depends on the spin. 
Matching the amplitudes of $\psi$ to leading order at $y=Y$
\be
{1 \over z}  H_\nu^{(2)}(z) {\Big |}_{z= \sqrt{2 \omega s}}
= \sigma \{ \alpha J_\nu(\sigma) + \beta Y_{\nu} (\sigma) \}
 {\Big |}_{\sigma= \sqrt{2 \omega s}},
\ee
one finds with $\beta=0$ that
\begin{eqnarray}
\nu = 1 &: &\alpha = { i \over \pi \omega^2 s^2}  \nonumber  \\
\nu = 2 &: &\alpha = { 4 i \over \pi \omega^3 s^3 } \\
\nu = 3 &: &\alpha = { 48 i \over \pi \omega^4 s^4}.  \nonumber
\end{eqnarray}

This means that we take only the normalizable modes and turn off 
the non-normalizable modes.
Now we calculate the flux. The asymptotic form of
the normalizable mode($y \to 0$) is
\be
\psi (\sigma) = \sigma \alpha J_\nu (\sigma) \simeq
\alpha \sqrt{{ \sigma \over 2 \pi}}
\exp \{ i ( \sigma - {1 \over 2} \nu \pi - {1 \over 4} \pi ) \}.
\ee
Then its flux, defined by
\be
F = {1 \over 2 i} \{ \psi^* { 1 \over y} \partial_y \psi -
               \psi { 1 \over y} \partial_y \psi^*  \},
\ee
is given by
\be
F_\infty = { \omega^2 |\alpha|^2 \over 2 \pi} .
\ee
Since the ingoing part of the wavefunction at horizon ($y \to \infty$)
 is given by
\be
\psi (z) = { 1 \over z} H^{(2)}_\nu (z) \simeq
\sqrt{ { 2 \over  \pi z^3} }
\exp \{- i ( z - {1 \over 2} \nu \pi - {1 \over 4} \pi ) \},
\ee
the ingoing flux at horizon is takes the form
\be
F_0 =  { 1 \over 2 \pi s^2}.
\ee
Taking the ratio of the flux across the horizon to the 
flux at infinity, we get the absorption probability
\be
{\cal A} = { F_0 \over F_\infty } =
{ 1 \over |\alpha|^2 \omega^2 s^2  }.
\ee
The absorption cross section is given by $\sigma_{abs} = {\cal A}/\omega$
in three dimensions.
Inserting $\alpha$ for $\nu = 1,2,3$, we have
\begin{eqnarray}
\nu = 1 &: &\sigma_{abs} = \pi^2 \omega s^2 \nonumber  \\
\nu = 2 &: &\sigma_{abs} = { \pi^2 \omega^3 s^4 \over 16} \label{abs-match0} \\
\nu = 3 &: &\sigma_{abs} = { \pi^2 \omega^5 s^6 \over {64 \times 36}}  \nonumber
\end{eqnarray}
Note that the Poincar\'e coordinates $(y,t,x)$ 
are dimensionless.  
By switching on $R$ we can recover the correct scale for cross section as 
\begin{eqnarray}
\nu = 1 &: &\sigma_{abs} = \pi^2 \omega R^2 s^2 \nonumber  \\
\nu = 2 &: &\sigma_{abs} = { \pi^2 \omega^3 s^4 R^4 \over 16} 
\label{abs-match} \\
\nu = 3 &: &\sigma_{abs} = { \pi^2 \omega^5 R^6 s^6 \over {64 \times 36}}.  \nonumber
\end{eqnarray}
The power of $R$ is in parallel with $s$ .

It is very important to compare our results (\ref{abs-match}) 
with those in the literature.  The
general formula for the greybody factor is given by up to the 
constant factor $C$ 
\begin{eqnarray}
\sigma^{h_R,h_L}_{abs} & = & C { (2 \pi R T_R)^{2 h_R - 1}
(2 \pi R T_L)^{2 h_L - 1} \over \omega \Gamma(2 h_R) \Gamma(2 h_L) }
    \sinh \Big ( {\omega \over 2 T_H} \Big ) \nonumber  \\
&~~~~~~ \times & \Big | \Gamma(h_R - i {\omega \over 4 \pi T_R})
     \Gamma(h_L - i {\omega \over 4 \pi T_L} ) \Big |^2,
\label{sigma-general}
\end{eqnarray}
where
\be
{2 \over T_H} = {1 \over T_L}+{1 \over T_R}
\ee
and $\Gamma$ is the gamma function.
Here $C$ is $2(h_R + h_L -1)^2 $ in the effective string method 
by Gubser\cite{Gub97PRD7854} and in 
the boundary CFT approach by Teo\cite{Teo98PLB269}.  Further 
$C=2 ( h_R + h_L -1)$ in the AdS$_3$-bulk calculation by 
Lee and Myung\cite{Lee9808002} and 
$C=(h_R + h_L)(h_R + h_L -1)$ in the bulk-boundary 
calculation by M\"uller-Kirstern, {\it et al.}\cite{Mul9809193}. 
The last three 
calculations are valid only for $h_L = h_R$. Gubser's calculation 
can accommodate the case of non-zero spin ($h_L \neq h_R$), 
where $h_L + h_R$ is an integer. 
But his result is still problematic in the case of $h_R=0$(or $h_L=0$).
Since our calculation is valid for the test fields with non-zero spin, we
will compare the two results.
For $\nu = 1$, we put $(h_R,h_L) = (0,2)$ in $\sigma^{h_R,h_L}_{abs}$
and compare the result with (\ref{abs-match}).
Note that $(2,0)$ and $(0,2)$ operators are coupled to 
gauge bosons in AdS$_3$.  The result is
\be
\sigma^{0,2}_{abs}= {2 \pi^2 R^2 \omega \over \Gamma(4) \Gamma(0) }
\left \{ 1 + \Big ( {2 \pi T_L \over \omega} \Big )^2 \right \}
{ \sinh \Big ( {\omega \over 2 T_H} \Big ) \over
  \sinh \Big ( {\omega \over 2 T_L} \Big )
  \sinh \Big ( {\omega \over 2 T_R} \Big ) }.
\label{abs-02}
\ee
Before we proceed, we note that $\sigma_{abs}^{h_R, h_L}$ is derived 
from the BTZ coordinates ($r, t, x^5$), while (\ref{abs-match}) is calculated 
with the Poincar\'e coordinates ($y,t,x$).
Hence, to compare our result of (\ref{abs-match}) with 
(\ref{abs-02}), one has to take the low temperature limit.
Taking this limit ($T_{L,R} \ll \omega<1 $), 
then one finds from (\ref{abs-02})
\be
\sigma^{0,2}_{abs} \to  {2 \pi^2 R^2 \over \Gamma(4) \Gamma(0)}
\omega,
\ee
which agrees with the leading term of (\ref{abs-match}) for $s=2, \nu = 1$
up to some factor. 
Although $\sigma^{0,2}_{abs}$ depends on $\omega$, it includes a 
singular term of $1/\Gamma(0)$. Hence the effective string 
approach does not give us the precise result 
for calculations of $\sigma^{0,2}_{abs}$, 
$\sigma^{0,3}_{abs}$, and $\sigma^{0,4}_{abs}$. 
We note here that our analysis for $s=2, \nu=1$ is unique because 
the gauge bosons are absent on the supergravity side.
Our method seems to be 
very useful for calculations with $h_R=0$(or $h_L=0$).
For $\nu = 2$, we can compare the result for $(h_R,h_L) = (1,2)$ and $(2,1)$
which are coupled to two intermediate scalars ($\eta, \xi$)
 in AdS$_3$.  For $(2,1)$, we have
\be
\sigma^{2,1}_{abs}= {8 \pi^2 R^4 \omega^3 \over \Gamma(4) \Gamma(2) }
\left \{ 1 + \Big ( {2 \pi T_L \over \omega} \Big )^2 \right \}
{ \sinh \Big ( {\omega \over 2 T_H} \Big )  \over
\sinh \Big ( {\omega \over 2 T_L} \Big )
\sinh \Big ( {\omega \over 2 T_R} \Big ) },
\ee
and similar one for $(1,2)$ by exchanging $T_L$ and $T_R$.
In the low temperature limit, one finds 
$\sigma^{2,1}_{abs} \propto \omega^3$.
We note that in this limit, the effective string calculation agrees 
with that of the AdS$_3$ bubble scattering for $(\eta, \xi)$\cite{Kle97NPB157}.
Also this agrees with (\ref{abs-match}) for $\nu =2$.
It is straightforward to show that the fixed scalar $\lambda$ of 
$\nu=3$ case in (\ref{abs-match}) 
agrees in the low temperature limit  of 
$\sigma_{abs}^{3,1}$ and $\sigma_{abs}^{1,3}$,
and with the AdS$_3$ bubble calculation\cite{Cal97NPB65}. 
The above confirms clearly that our result (\ref{abs-match}) is correct 
in the low temperature limit.

\section{First-order Matching}
As one would expect, the dominant corrections to the absorption cross
section may arise from the matching at the dual point
$y = Y = \sqrt{2s/\omega}$.  We will follow the approach of
\cite{Gub9803023,Tay9806132} to look for the field solution as power
series in $\omega$ and $s$.  In the region of $y < Y$ 
the right hand side of (\ref{fareq})
 is considered as a small correction to the zeroth-order solution.  However
as we approach the dual point $y = Y$, this term acts as a
correction to the zeroth-order solution of the same order.
In the near region of $y < Y$, we look for a
perturbative solution
$f_\nu(\sigma) = f_\nu^0(\sigma) + f_\nu^1(\sigma)$
where the zeroth-order solution is
\be
f_\nu^0(\sigma) = \alpha J_\nu(\sigma)
\ee
 and $f_\nu^1$ satisfies the inhomogeneous equation
\be
\sigma^2 f_\nu^{1\prime\prime} + \sigma f_\nu^{1\prime} + (\sigma^2 - \nu^2)
f_\nu^1 = - { 4 s^2 \sigma^2 \over \omega^2 R^{\prime 4}} f_\nu^0.
\ee
Solving this second-order equation for $f_\nu^1$, we have
\be
f_\nu^1(\sigma) = {\pi \over 2} \int^\sigma d x
{ 4 s^2 x \over \omega^2 R^{\prime 4}} f_\nu^0(x)
\{ J_\nu(x)Y_\nu(\sigma) - J_\nu(\sigma)Y_\nu(x) \}.
\ee
There is ambiguity in $f_\nu^1(\sigma)$ in the sense that  one can add
to it any solution of the homogeneous equation. We can fix this
ambiguity by demanding that the solutions of the near
and far regions match to order of $\omega s$ (or
$\omega s \ln(\omega s)$ for $\nu =1 $ case) in the transition region.
Substituting the form of $f_\nu^0(\sigma)$ and retaining the leading
terms in $\omega s$ ($\omega s \ln(\omega s)$), we find that the
first-order solution in the far region
$\psi = \sigma f_\nu(\sigma) = \sigma ( f_\nu^0(\sigma) + f_\nu^1(\sigma) )$
is given by
\begin{eqnarray}
\nu = 1 &: &\psi(\sigma) = {\alpha \over 2} \sigma^2
( 1 - {1 \over 8} \sigma^2
+ { s^2 \over 2 \omega^2 R^{\prime 4} } \sigma^2 ) \nonumber  \\
\nu = 2 &: &\psi(\sigma) = {\alpha \over 8} \sigma^3
 ( 1 - {1 \over 12} \sigma^2
+{1 \over 3}{ s^2 \over \omega^2 R^{\prime 4} } \sigma^2 )  \\
\nu = 3 &: &\psi(\sigma) = {\alpha \over 48} \alpha \sigma^4
 ( 1 - {1 \over 16} \sigma^2
+{1 \over 4}{ s^2 \over \omega^2 R^{\prime 4} } \sigma^2 )  \nonumber
\end{eqnarray}
We repeat the same procedure in the far region $y >Y$ to
find that the first-order solution in the transition region
$\psi = \phi(z)/ z = (\phi_0(z) + \phi_1(z))/z$ is given by
\begin{eqnarray}
\nu = 1 &: &\psi (z)= {2 i  \over \pi z^2}
\{ 1 - {1 \over 2} z^2 \ln z
+{\omega^2 s^2 \over 2 z^2} (1 + {3 \over 2} z^2 \ln z ) \} \nonumber  \\
\nu = 2 &: &\psi (z)= {4 i  \over \pi z^3}
( 1 + {1 \over 4} z^2 + {\omega^2 s^2 \over 3 z^2} )   
\label{leading-match} \\
\nu = 3 &: &\psi (z)= {16 i  \over \pi z^4}
( 1 + {1 \over 8} z^2 + {\omega^2 s^2 \over 4 z^2} ) \nonumber
\end{eqnarray}
Now we compare these two solutions at the matching point
$y=Y$. Since $\omega s$ is much smaller than one,
both $\sigma$ and $z$ are also small in the
matching region $\sigma = z =\sqrt{2 \omega s}$, and our
perturbative expansion is valid.  The mismatch between these two
solutions requires that one take
\begin{eqnarray}
\nu = 1 &: & \alpha = { i \over \pi \omega^2 s^2}
\{ 1 -  {1 \over 2} \omega s \ln (2 \omega s) \}      \nonumber  \\
\nu = 2 &: & \alpha = { 4 i \over \pi \omega^3 s^3}
\{ 1 + ( {5 \over 6} - {2 \over 3} {s^2 \over \omega^2 R^{\prime 4}} ) \omega s \}  \\
\nu = 3 &: & \alpha = { 48 i \over \pi \omega^4 s^4}
\{ 1 + ( {1 \over 2} - {1 \over 2} {s^2 \over \omega^2 R^{\prime 4}}) \omega s \}.
     \nonumber
\end{eqnarray}
This implies that the absorption cross sections behave as
\begin{eqnarray} \label{csour}
\nu = 1 &: &\sigma_{abs} = \pi^2 \omega R^2 s^2
\{ 1 + \omega s \ln (2 \omega s) \},
 \nonumber  \\
\nu = 2 &: &\sigma_{abs} = { \pi^2 \omega^3 R^4 s^4 \over 16}
\{ 1 - ( {5 \over 3} - {4 \over 3} {s^2 \over \omega^2 R^{\prime 4}} ) 
\omega s \},  \label{abs-cross}\\
\nu = 3 &: &\sigma_{abs} = { \pi^2 \omega^5 R^6 s^6 \over {64 \times 36}}
\{ 1 - ( 1 - {s^2 \over \omega^2 R^{\prime 4}}) \omega s \}  \nonumber
\end{eqnarray}
in the low temperature limit.

\section{Discussion}
From the AdS$_3$/CFT$_2$ correspondence($\nu=h_R+ h_L -1$), 
the $\nu=1$ case contains a 
minimally coupled scalar $\Phi$ which couples to (1,1) operator, and
the new gauge bosons to (2,0) and (0,2) operators\cite{Mal9804085}. 
The $\nu=2$ case involves a new field to (3/2, 3/2) operator, and  
two intermediate scalars ($\eta, \xi$) to (2,1), (1,2) 
operators\cite{Kle97NPB157}. 
The $\nu=3$ accommodates the dilaton (fixed scalar $\nu$) to (2,2), 
and the fixed scalar $\lambda$ to (2,2), (3,1), and (1,3) 
operators\cite{Cal97NPB65}. 
Taylor-Robinson showed that a minimally coupled scalar 
with $s=0, \nu=1$ receives 
logarithmic corrections in the cross section by semi-classical 
calculation, effective string approach, 
and AdS$_3$/CFT$_2$ correspondence\cite{Tay9806132}. 
Her geometry corresponds to an AdS$_3$ bubble(the near-horizon 
AdS$_3 \times$S$^3$ but 
with asymptotically flat space). Hence she needs a self-dual point 
which is the effective radius($R$) of the AdS$_3$ 
bubble for matching procedure. 
Here the self-dual point plays a role of the transition 
point from AdS$_3$ (near-horizon) to Minkowski space (asymptotically 
flat space). We remind the reader that our geometry is an exact AdS$_3$. Thus 
we don't need to introduce such a self-dual point that exactly plays 
the same role of $R$.  Instead, we introduce a spin-dependent 
matching point ($Y = \sqrt{2 s \over \omega}$) for computation. 

In this work we consider only the test fields with non-zero spin.
Unfortunately we do not thus obtain any information for a minimally 
coupled scalar with zero spin.
Improved matching in an exact AdS$_3$ for gauge bosons leads to a 
logarithmic correction to 
the absorption cross section. 
These gauge bosons appear in the resolution to the puzzle of the 
missing states between CFT$_2$ and
supergravity\cite{Vaf9804172,Boe9806104}.
Actually one cannot find these on the supergravity side.
In this sense they do not belong to physical fields.
But these are chiral primaries which correspond to the descendent 
of the identity operator in the CFT$_2$\cite{Mal9804085,Boe9806104}.
Here we can include these gauge bosons to study the exact AdS$_3$ 
scattering by considering the spin-dependent wave equation.
On the other hand, the $\nu=2,3$ cases including 
three physical fields($\eta, \xi, \lambda$) 
do not contain any logarithmic correction in the first-order approximation.

How do we interpret this result? First let us introduce the exact 
AdS/CFT lore which states that the conformal limit of gauge theory 
corresponds precisely with scattering from an exact AdS$_3$ background.
So there is no chance that the logarithmic term appears in an exact AdS$_3$ 
scattering. For example, this lore was proven partly 
for a minimally coupled scalar($\Phi$) and the dilaton($\nu$) 
in an exact AdS$_3$ scattering\cite{Lee9903054,Lee98PRD104013}.
Here we have proven this lore for three fields($\eta, \xi, \lambda$).
If we follow this lore, it is very hard to understand 
our logarithmic correction to the cross section for the 
gauge bosons.
Now we assume that the AdS/CFT lore is valid only for the 
physical test fields. Then there is 
a chance with logarithmic term in the exact AdS$_3$ scattering 
of the new gauge bosons, since these belong to unphysical fields.

In this direction we can relate the new gauge bosons 
to singletons in AdS$_5\times$S$^5$.
These are AdS$_5$ degrees of freedom that are pure gauge in the bulk 
and can be gauged away completely except at 
the boundary\cite{Kim85PRD389,Fer9807090}.
Here the relevant AdS$_5$ supergroup is SU(2,2$|$4).
For AdS$_3\times$S$^3$, the supergroup is given by SU(1,1$|$2) and thus 
the gauge bosons do not have an exact singleton representation.
But something similar happens in our case.
Let us consider SU$_R$(1,1$|$2)$\times$SU$_L$(1,1$|$2)
Chern-Simons theory. It does not have any propagating degree of 
freedom in an exact AdS$_3$ spacetime because it belongs to a topological 
field theory.
However the gauge field is subject to a boundary condition that contains 
fields living at the boundary.
These fields are generators of the right and left-moving chiral algebras.
Although the gauge bosons do not belong precisely to the 
singleton representation, these take 
similar properties as singletons\cite{Kog98PLB77}.
More recently, Kogan proposed a relationship between singletons in 
AdS and logarithmic conformal field theories on its 
boundary\cite{Kog9903162}.
He showed explicitly that the bulk AdS Lagrangian for a singleton 
dipole pair induces the two-point correlation function for logarithmic 
pair on the boundary.
This means that although we do not study the corresponding 
CFT$_2$ on the boundary, our logarithmic 
correction (\ref{abs-cross}) through (\ref{leading-match}) to the new 
gauge bosons may be understood in relation to singletons.
Consequently, our logarithmic correction represents that 
the new gauge bosons play the role of singletons.
This is also compatible with 
the AdS/CFT lore if this lore is suitable for the physical fields such as
a minimally coupled scalar($\Phi$), dilaton($\nu$), 
intermediate scalars($\eta, \xi$), and the fixed scalar($\lambda$).

\section*{Acknowledgement}
We would like to thank S.J. Rey for useful discussions.
This work was supported in part by the Basic Science Research Institute
Program, Minstry of Education, Project NOs. BSRI-98-2441 and
BSRI-98-2413 and grant from Inje University, 1998.


\begin{references}

\bibitem{Mal9711200} 
  J. Maldacena, Adv. Theor. Math. Phys. {\bf 2}, 231(1998), hep-th/9711200.
\bibitem{Gub9802109}
  S.S. Gubser, I.R. Klebanov and A.M. Polyakov, 
        Phys. Lett. {\bf B428}, 105(1998), hep-th/9802109.
\bibitem{Wit9802150}
  E. Witten, Adv. Theor. Math. Phys. {\bf 2}, 253(1998), hep-th/9802150.
\bibitem{Sfe98NPB179}
  K. Sfethos and K. Skenderis, Nucl. Phys. {\bf B517}, 179(1998), 
           hep-th/9711138;
  S. Hyun, hep-th/9704005.
\bibitem{Mal9804085}
  J. Maldacena and A. Strominger, hep-th/9804085.
\bibitem{Gub9803023} 
  S. Gubser, A. Hashimoto, I. Klebanov and M Krasnitz, 
          Nucl. Phys. {\bf B526}, 393(1998), hep-th/9803023.
\bibitem{Tay9806132} 
  M. Taylor-Robinson, hep-th/9806132.
\bibitem{Lee98PRD104013}
  H.W. Lee and Y.S. Myung, Phys. Rev. {\bf D58}, 104013(1998),
  hep-th/9804095;
  H.W. Lee, N.J. Kim, and Y.S. Myung, hep-th/9805050.
\bibitem{Bal99PRD046003}
  V. Balasubramanian, P. Kraus, and A. Lawrence, 
  Phys. Rev. {\bf D59}, 046003(1999), hep-th/9805171.
\bibitem{Dou9902022}
  M. Douglas and S. Randjbar-Daemi, hep-th/9902022.
\bibitem{Sus9901079}
  L. Susskind, hep-th/9901079; J. Polchinski, hep-th/9901076.
\bibitem{Bal9902052}
  V. Balasubramanian, P. Krauss and A. Lawrence, hep-th/9902052.
\bibitem{Gid9903048}
  S. Giddings, hep-th/9903048.
\bibitem{Lee9903054}
  H.W. Lee and Y.S. Myung, hep-th/9903054.
\bibitem{Vaf9804172}
  C. Vafa, hep-th/9804172.
\bibitem{Boe9806104}
  J. de Boer, hep-th/9806104.
\bibitem{Lee9808002}
  H.W. Lee and Y.S. Myung, hep-th/9808002.
\bibitem{Sat9810135}
  Y. Satoh, hep-th/9810135.
\bibitem{Gub97PRD7854} 
  S.S. Gubser, Phys.Rev.{\bf D56}(1997) 7854, hep-th/9706100.
\bibitem{Teo98PLB269}
  E. Teo, Phys. Lett. {\bf B436}, 269(1998), hep-th/9805014.
\bibitem{Mul9809193}
  H. M\"uller-Kirsten, N. Ohta, and J. Zhou, hep-th/9809193.
\bibitem{Kle97NPB157}
  I.R. Klebanov, A. Rajaraman and A. Tseytlin, Nucl. Phys.
            {\bf B503}, 157(1997), hep-th/9704112.
\bibitem{Cal97NPB65}
  C. Callan, S. Gubser, I.G. Klebanov and A. Tseytlin, Nucl. Phys.
    {\bf B489}, 65(1997), hep-th/9610172;
  M. Krasnitz and I.G. Klebanov, hep-th/9703216;
  H.W. Lee, Y.S. Myung and J.Y. Kim, Phys. Rev. {\bf D58},
     104006(1998), hep-th/9708099.
\bibitem{Bir97PLB281} 
  D.B. Birmingham, I. Sachs and S. Sen, Phys. Lett.  {\bf B413}, 281(1997), 
         hep-th/9707188;
  H.W. Lee, N.J. Kim, and Y.S. Myung, Phys. Lett. {\bf B441}, 83(1998),
  hep-th/9803227.
\bibitem{Abr66}
  M. Abramowitz and I. Stegun, {\it Handbook of Mathematical 
  Functions} (Academic Press, New York, 1966).
\bibitem{Kim85PRD389}
  H.J. Kim, L.J. Romans, and P. van Nieuwenhuizen, 
      Phys. Rev. {\bf D32}, 389(1985).
\bibitem{Fer9807090}
  S. Ferrara and A. Zaffaroni, hep-th/9807090.
\bibitem{Kog98PLB77}
  I.I. Kogan and A. Lewis, Phys. Lett. {\bf B432}, 77(1998), hep-th/9802102.
\bibitem{Kog9903162}
  I.I. Kogan, hep-th/9903162.

\end{references}
\end{document}